\documentclass[12pt]{article}  
\def\lsim{\mathrel{\lower2.5pt\vbox{\lineskip=0pt\baselineskip=0pt  
          \hbox{$<$}\hbox{$\sim$}}}}  
\def\gsim{\mathrel{\lower2.5pt\vbox{\lineskip=0pt\baselineskip=0pt  
          \hbox{$>$}\hbox{$\sim$}}}}  
\def\real{\mathrel{\lower.0pt \hbox{$I\!\!R$}}}  
  
\begin{document}     
\markright{Moduli fields and brane tensions:   
generalizing the junction conditions \hfil}  
\def\Barcelo{Barcel\'o}  
\def\Paris{Par\'\i{}s}  
\title{\Large   
\bf Moduli fields and brane tensions: \\  
\bf generalizing the junction conditions}  
\author{Carlos \Barcelo\ and Matt Visser\\[2mm]  
{\small \it   
Physics Department, Washington University,   
Saint~Louis, Missouri 63130-4899, USA.}}  
\date{{\small 2 August 2000; \LaTeX-ed \today}}  
\maketitle  
\begin{abstract}  
  
Taking the Randall--Sundrum models as background scenario, we derive 
generalized Israel--Lanczos--Sen thin-shell junction conditions for 
systems in which several bulk scalar fields are non-minimally coupled 
to gravity. We demonstrate that the form of the junction conditions 
(though arguably not the physics) depends on the choice of frame. We 
show that generally (in any frame except the Einstein frame) the 
presence of a thin shell induces discontinuities in the normal 
derivative of the scalar field, even in the absence of any direct 
interaction between the thin shell and the scalar field. For some 
exceptional scalar field configurations the discontinuities in the 
derivatives of the metric and the scalar fields can feed back into 
each other and so persist even in the absence of any thin shell of 
stress-energy.

\vspace*{5mm}  
\end{abstract}  
\vfill  
\hrule  
\bigskip  
\centerline{\underline{E-mail:} {\sf carlos@hbar.wustl.edu}}  
\centerline{\underline{E-mail:} {\sf visser@kiwi.wustl.edu}}  
\bigskip  
\centerline{\underline{Homepage:} {\sf http://www.physics.wustl.edu/\~{}carlos}}  
\centerline{\underline{Homepage:} {\sf http://www.physics.wustl.edu/\~{}visser}}  
\bigskip  
\centerline{\underline{Archive:}  
{\sf hep-th/0008008}}  
\bigskip  
\hrule  
\clearpage  
\def\Box{\nabla^2}  
\def\d{{\mathrm d}}  
\def\ie{{\em i.e.\/}}  
\def\eg{{\em e.g.\/}}  
\def\etc{{\em etc.\/}}  
\def\etal{{\em et al.\/}}  
\def\S{{\mathcal S}}  
\def\I{{\mathcal I}}  
\def\L{{\mathcal L}}  
\def\H{{\mathcal H}}  
\def\M{{\mathcal M}}  
\def\R{{\mathcal R}}  
\def\K{{\mathcal K}}  
\def\eff{{\mathrm{eff}}}  
\def\Newton{{\mathrm{Newton}}}  
\def\bulk{{\mathrm{bulk}}}  
\def\brane{{\mathrm{brane}}}  
\def\matter{{\mathrm{matter}}}  
\def\tr{{\mathrm{tr}}}  
\def\normal{{\mathrm{normal}}}  
\def\implies{\Rightarrow}  
\def\half{{1\over2}}  
\def\SIZE{1.00} 

\section{Introduction}  
\label{S:introduction}  
  
The gravitational interaction is by many orders of magnitude the  
weakest of all known interactions. The Planck length $l_P$, related to  
Newton's constant through $ l_P=\sqrt{G\hbar /c^3}$, is $10^{12}$  
smaller than the smallest length scale explored with standard-model  
physics, and is more than $10^{32}$ smaller than the length scales  
probed by direct gravitational experiments.  It is commonly believed  
that the classical geometrical picture with which we interpret  
large-scale physics must at some stage be abandoned as we approach  
smaller and smaller scales, but when and how this exactly happens is a  
matter of much debate.  In the usual scenario we might think of the  
Planck length as a fundamental cutoff scale and carry the  
4-dimensional geometric picture down to (say) some hundreds of Planck  
lengths, at which stage we would presumably enter the domain of pure  
quantum gravity.  
  
A radically different possibility which has recently attracted much 
attention is that there might already be observable deviations from 
this usual scenario at sub-millimetre distances. In the ``large extra 
dimensions'' scenario the geometry of spacetime is the product of our 
usual 4-dimensional spacetime times some relatively large but 
mathematically compact extra dimensions~\cite{large-compact}, giving 
rise to a rich phenomenology at low energies.  (At even smaller scales 
this geometrical multi-dimensional continuum would melt into a more 
fundamental quantum gravity theory, such as for instance string 
theory.)  In this scenario the Planck length would be an effective 
scale derived from the more fundamental electroweak scale, and the 
hierarchy problem (between these energy scales) would be translated 
into a hierarchy between the fundamental and the compactification 
scales.  In this approach, the standard-model fields in our 
4-dimensional hypersurface would not be part of a standard 
Kaluza--Klein reduction~\cite{large-compact}. (See also~\cite{old-kk} 
for a survey of other older non-standard Kaluza--Klein proposals.) 
    
Another possibility with similar phenomenological richness which,  
moreover, avoids the new hierarchy problem left over by the previous  
approach, is the ``brane world'' scenario. Randall and Sundrum  
recently proposed that the extra dimensions could be non-compact  
provided that they were sufficiently ``warped''~\cite{RS}.  (For  
similar suggestions, see also~\cite{Gogberashvili}  
and~\cite{arkani-et-al}. Considerably older precursors to the idea of  
non-compact extra dimensions can be found in~\cite{old-non-compact}  
and~\cite{Gell-Mann}.)  Inspired by string theory, Randall and Sundrum  
proposed than we live on a 3-brane [(3+1)-dimensional hypersurface]  
evolving in a (4+1)-dimensional anti-de Sitter bulk.  Everything is  
confined to live on the brane, except for gravity itself.  To ``weakly  
trap'' gravity the specific form of the spacetime geometry provides a  
normalizable Kaluza--Klein zero mode, corresponding to a 4-dimensional  
graviton (trapped near but not on the brane), plus a continuum of  
massive Kaluza--Klein graviton modes modifying the gravitational  
interaction at small distance scales (scales comparable with the warp  
length).

Within this brane world scenario much work has been done analyzing the  
effective gravity induced on the  
brane~\cite{RS-cosmology,barcelo,RS-gravity}.  {From} a cosmological  
perspective, it has been demonstrated that we can recover a  
Friedmann-like evolution for the late brane world if there is a large  
negative cosmological constant in the bulk and a large positive  
intrinsic tension on the brane. These are carefully arranged to  
counteract each other, and so leave a small (possibly zero) effective  
cosmological constant~\cite{RS-cosmology,barcelo}.  In the early brane  
world, when the energy density of matter fields reaches values of the  
order of the brane tension, the cosmological evolution will be  
modified giving place to a polynomial inflation~\cite{RS-cosmology}.  
{From} a more general point of view, the circumstances under which it  
is possible to recover standard general relativity on the brane have  
been studied, and possible modifications to the standard picture have  
been analyzed~\cite{RS-gravity}.  
  
It is important to realize that apart from gravity itself there could  
be other fields not constrained to live on the brane. By considering  
the low-energy field theory limit of string theory one can deduce the  
existence of a plethora of bulk fields: primarily the dilaton field  
and the torsion field, but generically there will also be many moduli  
fields (corresponding to ``soggy'' deformations of the six-dimensional  
Calabi--Yau manifold that is also part of these string-induced  
scenarios). Also, adopting a modification of more conventional  
Kaluza--Klein ideas, it is possible that our 4-dimensional (light)  
standard-model fields might be the reduction to 4 dimensions of some  
5-dimensional fields with Planck scale masses, with the effective  
4-dimensional mass being greatly suppressed compared to the more  
fundamental 5-dimensional mass parameter~\cite{goldberger}. Thus it is  
important to understand the role that additional bulk fields might  
play in the brane world scenario.  
  
For instance, the existence of scalar fields in the bulk could provide  
an answer to the cosmological constant problem~\cite{lambdaprob} (for  
early ideas on this subject, see~\cite{shaposhnikov}). For arbitrary  
values of the brane tension (an intrinsic vacuum energy density) there  
exist solutions of gravity plus a dilaton scalar field which lead to  
zero bulk cosmological constant, and are compatible with a  
4-dimensional Poincare invariant brane world~\cite{lambdaprob}. (This  
implies a vanishing effective cosmological constant for the brane  
evolution.)  In these solutions the warp factor becomes zero at a  
finite proper distance from the brane in the extra dimension and so it  
is ``effectively compactified'' leaving standard 4-dimensional gravity  
for large scales.  
  
In the models considered to date the bulk scalar fields have no direct 
coupling to the bulk scalar curvature. In the case of the dilaton 
field this is accomplished by working in the Einstein frame and not in 
the (perhaps more natural) string frame.  There are many interesting 
scalar fields theories that cannot easily be dealt with in the current 
approach, such as the Brans--Dicke theory (and more generically, 
scalar-tensor theories), non-minimally coupled scalars, or the dilaton 
field in the string frame. In this paper we will work out a formalism 
to deal with a large class of scalar field theories in arbitrary 
frames, by allowing arbitrary couplings to the bulk curvature scalar. 
  
In section \ref{S:action}, we will review the Israel--Lanczos--Sen 
thin-shell formalism~\cite{ILS}, obtaining the junction conditions in 
an easy to generalize manner by beginning directly from the 
gravitational action.  This method will be then be applied the case in 
which several very general scalar fields interact with gravity. We 
will obtain generalized junction conditions for these systems. In 
sections {\ref{S:exception}} and {\ref{S:moduli}} we will discuss these 
new junction conditions. We will see that these junction conditions 
are now much more complicated than the usual ones, and that many 
intuitive arguments involving brane tensions no longer work.  The 
point is that the insertion of a ``kink'' in the geometry, achieved by 
inserting a thin shell of stress-energy, now gives a contribution to 
the formation of a ``kink'' in the scalar field configuration. 
{\emph{Vice-versa}}, the presence of a ``kink'' in the scalar field 
now makes a contribution to the formation of a ``kink'' in the 
geometry.  Finally, in section {\ref{S:summary}} we summarize our 
results and try to put things in perspective.

\section{Action principle and junction conditions}  
\label{S:action}  
  
We are interested in geometries in $n$ spacetime dimensions that have 
an internal hypersurface of co-dimension one on which there is no well 
defined notion of tangent plane.  The derivatives of the metric, with 
respect to a proper normal coordinate across the hypersurface, undergo 
a jump on passing through it. We can interpret the presence of this 
jump in the normal derivative of the metric as produced by an 
infinitely thin shell of stress-energy located on the hypersurface 
(one can speak equivalently of a thin shell of stress-energy, an 
idealized domain wall, or an evolving $n-2$ brane).  The junction 
conditions give a relation between the intrinsic energy-momentum 
tensor of the brane and the form and strength of the jump in 
derivatives of the metric (encoded in the jump in the extrinsic 
curvature of the hypersurface as seen from its two faces).

\subsection{Israel--Lanczos--Sen junction conditions: Einstein Frame}  

Let us first review (and extend) the derivation of the standard  
Israel--Lanczos--Sen junction conditions.  We will use a method easily  
generalizable to more complicated systems involving non-minimal  
couplings between gravity and additional fields.  
(The paper by Chamblin and Reall~\cite{chamblin} contains a similar  
derivation of the Einstein frame junction conditions.)  
We start with the Einstein--Hilbert action, supplemented by the  
Gibbons--Hawking boundary term~\cite{gibbons}, and add both bulk  
fields and an intrinsic action for the brane:  
\begin{eqnarray}  
\S &=&   
{1 \over 2}   
\int_{{\mathrm{int}}(\M)} \sqrt{-g} \;\d^{n} x \;   
\left( R - 2 \Lambda \right)  
-  
\int_{\partial\M} \sqrt{-q} \;\d^{n-1}x \; K   
\nonumber\\  
&&  
+  
\int_{{\mathrm{int}}(\M)} \sqrt{-g} \;\d^{n} x \;  \L_\bulk  
+  
\int_{\rm brane} \sqrt{-q} \;\d^{n-1}x \; \L_\brane.  
\label{E:gravact}  
\end{eqnarray}  
Here $q_{AB}$ represents the induced metric, either on the  
$(n-1)$-dimensional hypersurface forming the boundary, or on that  
defined by the brane.  (We adopt the conventions of Misner, Thorne,  
and Wheeler~\cite{MTW}.) This expression makes perfectly good sense  
because the scalar of curvature is well defined (in the distributional  
sense) even at points on the brane itself.  
  
For definiteness we shall split the bulk matter Lagrangian $\L_\bulk$  
into contributions from the scalar moduli fields $\phi^i$, which for  
generality will be described by a non-linear sigma model, and other  
generic bulk fields denoted $\psi$. That is  
\begin{equation}  
\L_\bulk(g_{AB},\phi,\psi) =  
-{1\over2} H_{ij}(\phi) \;   
\left[ g^{AB} \; \partial_{A}\phi^i \; \partial_{B}\phi^j \right]  
- V(\phi,\psi) + \L_\bulk(g_{AB},\psi).  
\end{equation}  
As usual, we assume that $\L_\bulk(g_{AB},\psi)$ contains no second  
derivatives of $\psi$, and for convenience assume it contains not even  
first derivatives of the metric. (This last condition is not  
essential, but relaxing it would involve additional unnecessary  
algebraic complexity.)  
  
In the spirit of the brane world models, let us now deal with  
spacetimes that are decomposable as $\Sigma\times\real$, where $\real$  
represents a spatial dimension parametrized by $\eta$ and $\Sigma$ is  
a $n-1$ dimensional spacetime (locally, this is always possible). We  
consider a formal evolution in the $\eta$ parameter between two fixed  
hypersurfaces located at $\eta_i$ and $\eta_f$ (they are assumed to  
comprise the boundary $\partial\M$), leaving the brane located  
at $\eta=0$. (In most of the literature, the fifth dimension $\eta$ is  
taken to be spacelike, and we also take this option; it is for this  
reason that we speak of $\eta$ as ``formal'' evolution parameter, for  
timelike fifth dimension a few minus signs will change.)  The total  
action (\ref{E:gravact}) can be re-written as  
\begin{eqnarray}  
\S=  
\lim_{0^{-},0^{+}\rightarrow 0}\Bigg\{  
&&\hspace{-6mm}  
{1 \over 2} \int_{\eta_i}^{0^{-}}\sqrt{-g} \;\d\eta\;\d^{n-1} x   
\left[{}^{(n-1)}R -K^{AB}K_{AB}+ K^2 - 2\Lambda \right]  
\nonumber \\  
&&\hspace{-6mm}  
+{1 \over 2} \int^{\eta_f}_{0^{+}}\sqrt{-g} \;\d\eta\;\d^{n-1} x   
\left[{}^{(n-1)}R -K^{AB}K_{AB}+ K^2 - 2 \Lambda \right]   
\nonumber \\  
&&\hspace{-6mm}  
+{1 \over 2} \int_{0^{-}}^{0^{+}}\sqrt{-g} \;\d\eta\;\d^{n-1} x \;R  
-\int \sqrt{-q} \;\d^{n-1}x \; K \bigg|_{0^{-}}^{0^{+}}\Bigg\}  
\nonumber \\  
&&\hspace{-20mm}  
+  
\int_{{\mathrm{int}}(\M)} \sqrt{-g} \;\d^{n} x \;  \L_\bulk  
\nonumber \\  
&&\hspace{-20mm}  
+\int_{\rm brane} \sqrt{-q} \;\d^{n-1}x \;\L_\brane.  
\label{E:action}  
\end{eqnarray}    
In obtaining this expression we have used the fact that one can  
unambiguously separate $R$ into a term that contains no second  
derivatives of the metric, and a second term (a total divergence)  
isolating all the second derivatives of the metric:  
\begin{equation}  
R= {}^{(n-1)}R - K^{AB}\; K_{AB}+ K^2 + 2\nabla_A(n^A K),  
\label{E:r}  
\end{equation}  
with $n_A$ denoting the normal to the foliation.  To verify this, see 
for example, MTW (21.84), (21.86), and (21.88), supplemented with 
exercise (21.10), and equation (21.82), being careful to keep to a 
timelike hypersurface (spacelike normal) throughout~\cite{MTW}. Also 
note that in MTW conventions 
\begin{equation}  
K_{AB} = - {1\over2} \; {\partial q_{AB}\over\partial\eta}  
\end{equation}  
for both spacelike and timelike hypersurfaces.  
  
For the class of geometries we are considering the scalar of  
curvature is everywhere finite except for a possible delta-function  
contribution at $\eta=0$. If $K$ undergoes a finite jump at $\eta=0$,  
then the term with second derivatives of the metric (first derivatives  
of $K$) in equation (\ref{E:r}) will give rise to this delta-function  
contribution.  Then, it is easy to see that  
\begin{equation}  
\lim_{0^{-},0^{+}\rightarrow 0}\left[  
{1 \over 2} \int_{0^{-}}^{0^{+}}\sqrt{-g} \;\d\eta\;\d^{n-1} x \;R  
-  
\int \sqrt{-q} \;\d^{n-1}x \; K \bigg|_{0^{-}}^{0^{+}}\right]=0,  
\label{E:defint}  
\end{equation}  
independently of the existence or not of any delta-function  
contribution.  
  
As a consequence, the total action (though not the bulk  
Einstein--Hilbert Lagrangian) is finally formed by two standard bulk  
pieces [the first two terms in (\ref{E:action})] plus the contribution  
of the brane [last term in (\ref{E:action})]. We explicitly see that the  
action (\ref{E:gravact}) contains no second derivatives of the metric:  
\begin{eqnarray}  
\S  
&=&
{1 \over 2} \int_{\eta_i}^{0}\sqrt{-g} \;\d\eta\;\d^{n-1} x   
\left[{}^{(n-1)}R -K^{AB}K_{AB}+ K^2 \right]  
\nonumber \\  
&+&
{1 \over 2} \int^{\eta_f}_{0}\sqrt{-g} \;\d\eta\;\d^{n-1} x   
\left[{}^{(n-1)}R -K^{AB}K_{AB}+ K^2 \right]   
\nonumber \\  
&+&
\int_{{\mathrm{int}}(\M)} \sqrt{-g} \;\d^{n} x \;  \L_\bulk  
\nonumber \\  
&+&
\int_{\brane} \sqrt{-q} \;\d^{n-1}x \;\L_\brane.  
\label{E:action2}  
\end{eqnarray}    
This action can now be re-expressed in a Hamiltonian form  
\begin{eqnarray}  
\S&=&
\int_{\eta_i}^{0}\sqrt{-g} \;\d\eta\;\d^{n-1} x  
\left[\pi^{AB}\;{\d q_{AB}\over \d\eta}-\H(q_{AB},\pi^{AB})\right]  
\nonumber \\  
&+&
\int^{\eta_f}_{0}\sqrt{-g} \;\d\eta\;\d^{n-1} x  
\left[\pi^{AB}\;{\d q_{AB}\over \d\eta}-\H(q_{AB},\pi^{AB})\right]  
\nonumber\\  
&+&
\int_{\brane} \sqrt{-q} \;\d^{n-1}x\; \L_\brane,  
\end{eqnarray}  
where we have temporarily suppressed the bulk fields $\phi^i$ and  
$\psi$. Here $\H$ is the appropriate Hamiltonian density (its exact  
form is not needed for the following argument), and we use the  
following definition for the canonical momentum  
\begin{equation}  
\pi^{AB}={1 \over 2} \sqrt{-g} \left(K^{AB}-q^{AB}\;K\right).  
\label{E:momentum}  
\end{equation}  
The canonical Hamiltonian will also depend on the lapse and shift  
function of the standard $[(n-1)+1]$ ADM decomposition, but we have  
omitted it for simplicity (as it does not contribute in the following  
argument). We have partially fixed the gauge by placing the brane at  
$\eta=0$. Therefore, we are going to consider only diffeomorphisms  
that do not affect the boundaries or the location of the brane.  
  
By varying this action with respect $q_{AB}$ and $ \pi^{AB}$ we obtain   
\begin{eqnarray}  
\delta\S=&&\hspace{-6mm}  
\int_{\eta_i}^{0}\sqrt{-g} \; \d\eta\;\d^{n-1} x  
\left[{dq_{AB}\over d\eta} -{\partial \H \over \partial \pi^{AB} }\right]  
\delta \pi^{AB}  
\nonumber \\  
&&\hspace{-6mm}  
-\int^{\eta_f}_{0}\sqrt{-g} \;\d\eta\;\d^{n-1} x  
\left[{d\pi^{AB}\over d\eta} +{\partial \H \over \partial q_{AB} }\right]  
\delta q_{AB}  
\nonumber \\  
&&\hspace{-6mm}  
+\int \sqrt{-q} \;\d^{n-1}x \;\pi^{AB} \; \delta q_{AB}\bigg|_{\eta_i}^{\eta_f}  
\nonumber \\  
&&\hspace{-6mm}  
+\int \sqrt{-q} \;\d^{n-1}x \; (\pi^{-AB}-\pi^{+AB}) \; \delta q_{AB}\bigg|_{\eta=0}  
\nonumber \\  
&&\hspace{-6mm}  
+{1 \over 2}\int \sqrt{-q} \;\d^{n-1}x \; S^{AB} \; \delta q_{AB},  
\end{eqnarray}  
where $S^{AB}$ is the energy-momentum tensor of the brane.  
  
This action has an extremum among all the geometries that connect the  
two fixed hypersurfaces at $\eta_i$ and $\eta_f$ when: ({\em i}\,) it  
satisfies the classical equation of motion in the two bulk regions  
(the full diffeomorphism invariance in each bulk region will result in  
the additional classical constraint equations not of direct interest),  
and, ({\em ii}\,) when the junction condition  
\begin{equation}  
\pi^{+}_{AB}- \pi^{-}_{AB} = {1\over2}  \sqrt{-q} \; S_{AB}  
\label{E:pi+pi-}  
\end{equation}  
holds.

Now define   
\begin{equation}   
\K_{AB}\equiv  
K^{+}_{AB}-K^{-}_{AB},   
\label{E:israel-def}  
\end{equation}  
and substitute the explicit form of the momentum (\ref{E:momentum}) into  
(\ref{E:pi+pi-}).  We obtain the usual Israel--Lanczos--Sen junction  
conditions  
\begin{equation}   
\K_{AB} - \K \; q_{AB} =  
S_{AB}.   
\label{E:israel}  
\end{equation}  
For future generalization, we find it convenient to split this result  
into trace-free part and a trace, so that  
\begin{equation}   
\K_{AB} - {1\over n-1} \; \K \; q_{AB} =  
S_{AB} - {1\over n-1} \; S \; q_{AB},   
\label{E:israel-tracefree}  
\end{equation}  
and  
\begin{equation}   
\K= - {S \over n-2}.   
\label{E:israel-trace}  
\end{equation}  
This can be reassembled to yield the equivalent form  
\begin{equation}   
\K_{AB} =  
S_{AB}- {1 \over n-2 } \; S \;q_{AB}.  
\label{E:israel-reassemble}  
\end{equation}  
These junction conditions have a tremendous number of applications,  
well beyond the confines of the brane world scenario, and are a basic  
tool of general applicability~\cite{applications}.   
  
Now reinstate the non-gravitational bulk fields $\phi^i$ and $\psi$,  
and repeat the argument: there will be several additional conjugate  
momenta  
\begin{equation}  
\pi_{\phi^i} = - \sqrt{-g} \; H_{ij}(\phi) \;  
\left({\partial \phi^j\over\partial\eta}\right),  
\end{equation}  
and  
\begin{equation}  
\pi_{\psi} = \sqrt{-g}  \;  
\left(  
{\partial \L_\bulk(g_{AB},\psi)\over\partial(\partial\psi/\partial\eta)}  
\right).
\label{psiccm}    
\end{equation}  
There will also be additional junction conditions on the bulk fields  
$\phi^i$ and $\psi$:  
\begin{equation}  
(\pi_{\phi^i})^+ - (\pi_{\phi^i})^- =   
\sqrt{-q} \; {\partial\L_\brane\over\partial\phi^i},   
\end{equation}  
and  
\begin{equation}  
(\pi_{\psi})^+ - (\pi_{\psi})^- =   
\sqrt{-q} \; {\partial\L_\brane\over\partial\psi}.
\end{equation}  
Because we have decided on an explicit form for the $\phi^i$  
Lagrangian, we can simplify that junction condition  
considerably. First introduce the notation  
\begin{equation}  
J^i\equiv n^A (\partial_A \phi^i)^{+} - n^A (\partial_A \phi^i)^{-}   
\equiv   
\left({\partial\phi^i\over\partial\eta}\right)^+   
-   
\left({\partial\phi^i\over\partial\eta}\right)^-,  
\end{equation}  
Furthermore, let $H^{ij}(\phi)$ denote the inverse matrix to $H_{ij}(\phi)$.  
Then  
\begin{equation}  
J^i = -H^{ij}(\phi)\;{\partial \L_\brane \over \partial \phi^j}.   
\label{E:j-einstein}  
\end{equation}  
On the other hand, because we have made no specific commitment as to  
the form of $\L_\bulk(g_{AB},\psi)$ the best we can say for those  
fields is that  
\begin{equation}  
\left[  
{\partial \L_\bulk(g_{AB},\psi)\over\partial(\partial\psi/\partial\eta)}  
\right]^{+}
-
\left[  
{\partial \L_\bulk(g_{AB},\psi)\over\partial(\partial\psi/\partial\eta)}  
\right]^{-}
=   
{\partial\L_\brane(g_{AB},\psi)\over\partial\psi}.  
\label{genericjc}
\end{equation}  
Though abstract, this bulk-field junction condition is perhaps the  
clearest general statement one can make regarding the coupling of  
non-gravitational bulk fields to the brane.  In the next subsection we  
will see how this method of obtaining the junction conditions is  
easily generalizable to even more complicated systems, and in  
particular how to deal with generic frame dependence and non-minimal  
curvature coupling.  
  
\subsection{Moduli fields --- Generic frames}  
  
We are now going to analyze a system in which there is not only  
gravity in the bulk, but also additional bulk scalar fields with a  
{\emph{non-minimal coupling}} to the curvature scalar. Explicitly, we  
take an action of the form  
\begin{eqnarray}  
\S=&&\hspace{-6mm}  
{1 \over 2}\int_{{\rm int}(\M)} \sqrt{-g} \;\d^n x \;  
F(\phi) \; \left[ R -2\Lambda \right]   
-\int_{\partial\M} \sqrt{-q} \;\d^{n-1}x \; F(\phi)\;K  
\nonumber\\    
&&\hspace{-6mm}  
+\int_{{\rm int}(\M)} \sqrt{-g} \;\d^n x \;  
\left\{ -{1 \over 2}   
H_{ij}(\phi) \;   
\left[ g^{AB} \; \partial_{A}\phi^i \; \partial_{B}\phi^j \right]   
-V(\phi,\psi)   
+\L_\bulk(g_{AB},\psi)  
\right\}  
\nonumber \\    
&&\hspace{-6mm}  
+\int_{\rm brane} \sqrt{-q} \;\d^{n-1}x\;\L_\brane(q_{AB},\phi,\psi).  
\label{E:system}  
\end{eqnarray}  
The Gibbons--Hawking boundary term now takes the form    
\begin{equation}  
-\int_{\partial\M} \sqrt{-q} \;\d^{n-1}x \; F(\phi)\;K.  
\end{equation}  
This is specifically chosen to cancel the second derivatives of the  
metric in the Einstein--Hilbert action. The two (in principle)  
arbitrary functions, $F(\phi)$ and $H_{ij}(\phi)$, allow us to take into  
account a large class of possible moduli fields.  
  
In this notation the dilaton field in the string frame corresponds to  
\begin{equation}  
F(\phi)=e^{-2\phi}; \qquad H(\phi)=-4 \; e^{-2\phi}.  
\end{equation}  
In the Einstein frame the dilaton corresponds to  
\begin{equation}  
F(\phi)=1; \qquad H(\phi)={4\over n-2}.  
\end{equation}   
%
%
Standard non-minimally coupled scalars correspond to  
\begin{equation}  
F(\phi)=1-\xi_{ij}\;\phi^i\;\phi^j; \qquad   
H_{ij}(\phi)=\delta_{ij}.  
\end{equation}  
This is typically (though perhaps misleadingly) referred to as the  
Jordan frame, in this context meaning the frame in which these  
particular scalars have canonical kinetic energy.  
Note that non-minimally coupled scalars (\eg, conformally coupled) are 
from a quantum field theory point of view generic~\cite{scalars}; 
however from a gravitational point of view they are potentially 
problematic, leading to classical violations of the energy conditions 
and worse~\cite{scalars,Flanagan}. 
Note that you can always make non-minimal coupling go away by  
performing a conformal transformation and shifting frame to the  
Einstein frame --- there is however a conservation of difficulty and  
the price paid is to modify the scalar kinetic energies. It is often  
convenient to keep the scalar kinetic energies simple and instead  
confront non-minimal coupling head-on.  Finally, we can describe a  
Brans--Dicke theory by means of  
\begin{equation}  
F(\phi)=\phi; \qquad H(\phi)={\omega(\phi)\over\phi}.  
\end{equation}   
%
%
Confusingly enough, this is {\em also} typically referred to as the  
Jordan frame, in this context meaning the frame in which the kinetic  
energies of {\em all other} non-Brans--Dicke bulk fields are canonical.  
  
We will shortly see that the presence of a function $F(\phi)\neq 1$  
can yield to very interesting effects.  To also take into account  
possible interactions between the brane and the bulk fields, in our  
analysis we will allow the brane Lagrangian to depend arbitrarily on  
the metric and on the bulk scalar fields, (this is in addition to its  
dependence on the fields trapped on the brane).  We will  
allow also a bulk potential term for the scalar fields, and a  
cosmological constant in the bulk.    
  
Now, let us follow the same steps as in the previous subsection.  The  
whole action (\ref{E:system}) can be separated into two bulk  
contributions without any second derivative of the metric plus the  
brane action. {From} the term  
\begin{equation}  
{1 \over 2}\int \sqrt{-g} \; \d^n x \; F(\phi) \;  R  
\end{equation}  
we can see that the bulk actions include a term of the form  
\begin{equation}  
{1 \over 2} \int\sqrt{-g}\;\d\eta \;\d^{n-1} x \; F(\phi)  
\left[  
{}^{(n-1)}R -K^{AB}K_{AB}+ K^2   
-2{\partial\ln F(\phi)\over\partial\phi^i} \;  
\left({\partial\phi^i\over\partial\eta}\right) \; K  
\right],  
\end{equation}  
with the prime denoting ordinary derivative with respect to $\phi$.  
Indeed the total action is  
\begin{eqnarray}  
\S  
&=&
{1 \over 2} \int_{\eta_i}^{0}\sqrt{-g} \;\d\eta\;\d^{n-1} x   
\nonumber\\  
&&\qquad  
\; F(\phi)  
\left[  
{}^{(n-1)}R -K^{AB}K_{AB}+ K^2   
-2{\partial\ln F(\phi)\over\partial\phi^i} \;  
\left({\partial\phi^i\over\partial\eta}\right) \; K  
-2 \Lambda  
\right]  
\nonumber \\  
&+&
{1 \over 2} \int^{\eta_f}_{0}\sqrt{-g} \;\d\eta\;\d^{n-1} x   
\nonumber\\  
&&\qquad  
\; F(\phi)  
\left[  
{}^{(n-1)}R -K^{AB}K_{AB}+ K^2   
-2{\partial\ln F(\phi)\over\partial\phi^i} \;  
\left({\partial\phi^i\over\partial\eta}\right) \; K  
-2 \Lambda  
\right]  
\nonumber \\  
&+&
\int_{\eta_i}^{0}\sqrt{-g} \;\d\eta\;\d^{n-1} x   
\left\{   
- {1 \over 2}  H_{ij} (\phi) \; \left[  
g^{AB} \; \partial_{A}\phi^i \; \partial_{B}\phi^j    
\right]   
- V(\phi,\psi)  
+ \L_\bulk(g_{AB},\psi)  
\right\}  
\nonumber \\    
&+&
\int^{\eta_f}_{0}\sqrt{-g} \;\d\eta\;\d^{n-1} x   
\left\{   
- {1 \over 2} H_{ij}(\phi) \; \left[  
g^{AB} \; \partial_{A}\phi^i \; \partial_{B}\phi^j   
\right]   
- V(\phi,\psi)  
+ \L_\bulk(g_{AB},\psi)  
\right\}  
\nonumber \\    
&+&
\int_{\rm brane} \sqrt{-q} \;\d^{n-1}x \;\L_\brane.  
\label{E:action2a}  
\end{eqnarray}    
When re-expressed in a Hamiltonian form we get  
\begin{eqnarray}  
S&=&
\int_{\eta_i}^{0}\sqrt{-g} \;\d\eta\;\d^{n-1} x  
\left[  
\pi^{AB}\;{\d q_{AB}\over \d\eta}  
+\pi_{\phi^i}\;{\d \phi^i\over \d\eta}	  
+\pi_\psi \; {\d \psi\over \d\eta}  
-\H(q_{AB},\phi,\psi,\pi^{AB},\pi_\phi,\pi_\psi)  
\right]  
\nonumber \\  
&+&
\int^{\eta_f}_{0}\sqrt{-g} \;\d\eta\;\d^{n-1} x  
\left[  
\pi^{AB}\;{\d q_{AB}\over \d\eta}  
+\pi_{\phi^i}\;{\d \phi^i\over \d\eta}	  
+\pi_\psi \; {\d \psi\over \d\eta}  
-\H(q_{AB},\phi,\psi,\pi^{AB},\pi_\phi,\pi_\psi)  
\right]  
\nonumber \\  
&+&
\int_{\rm brane} \sqrt{-q} \;\d^{n-1}x\; \L_\brane,  
\end{eqnarray}  
where again we do not need to know the detailed form of the  
Hamiltonian.  Following the same reasoning as before, we conclude that  
the modified junction conditions read  
\begin{eqnarray}  
\pi^{+}_{AB}- \pi^{-}_{AB}  
=  
{1 \over 2}  \sqrt{-q} \; S_{AB},   
\label{E:ilsphi1}  
\\   
\pi^{+}_{\phi^i}-\pi^{-}_{\phi^i}  
=   
\sqrt{-q} \;   
{\partial \L_\brane \over \partial \phi^i},  
\label{E:ilsphi2}  
\\   
\pi^{+}_{\psi}-\pi^{-}_{\psi}  
=   
\sqrt{-q} \;   
{\partial \L_\brane \over \partial \psi},  
\label{E:ilsphi3}  
\end{eqnarray}  
but now with a different definition for the gravitational momentum  
\begin{equation}  
\pi_{AB}  
=  
{1 \over 2} \sqrt{-q} \; F(\phi)\; \left(K_{AB}-q_{AB}K\right)  
+{1 \over 2} \sqrt{-q} \; F'_i(\phi)\;   
\left({\partial\phi^i\over\partial\eta}\right) q_{AB}.  
\label{E:pigphi}   
\end{equation}  
Here for convenience we have defined  
\begin{equation}  
 F'_i(\phi) \equiv \partial_i F(\phi)   
\equiv {\partial F(\phi)\over\partial\phi^i}.  
\end{equation}  
The momentum canonically conjugate to $\phi$ is also altered and now reads  
\begin{equation}  
\pi_{\phi^i}  
=  
-\sqrt{-q}\;H_{ij}(\phi)\; \left({\partial\phi^j\over\partial\eta}\right)   
-\sqrt{-q}\; F'_i(\phi)\; K.  
\label{E:piphi}  
\end{equation}  
Obviously, the generic expression (\ref{psiccm}) for the momentum 
canonically conjugated to $\psi$ remains unchanged and so its 
associated junction condition (\ref{genericjc}). For simplicity, from 
here on we will not consider explicitly these additional fields.
Rearranging the previous expressions [(\ref{E:ilsphi1}) to  
(\ref{E:piphi})] and using the previous definition of $J^i$ we arrive at  
\begin{equation}  
F(\phi) \; \left(\K_{AB} - q_{AB} \; \K\right)   
+ F'_i(\phi) \; J^i \; q_{AB}   
=  S_{AB}.  
\label{E:j1}  
\end{equation}  
and  
\begin{equation}  
-H_{ij}(\phi) \; J^j - F'_i(\phi) \; \K =   
{\partial \L_\brane \over \partial \phi^i}.  
\label{E:j2}  
\end{equation}  
This shows how discontinuities in the extrinsic curvature mix with  
discontinuities in the normal derivatives of the scalar field. (The  
junction condition for $\psi$ is unaltered because of our technical  
assumption that $\L_\bulk(g_{AB},\phi)$ contains no metric  
derivatives.) In the Einstein frame [$F(\phi)=1$] this mixing switches  
off and we recover the ordinary Israel--Lanczos--Sen junction  
condition, supplemented by the bulk-field junction conditions of the  
preceding subsection.  If we move away from the Einstein frame (and  
it is often convenient to do so) the price paid is that the junction  
conditions become inextricably intertwined.  
  
For these generalized junction conditions it is extremely useful to  
separate the equation (\ref{E:j1}) into a trace-free portion and a  
trace. That is  
\begin{equation}   
\K_{AB} - {1\over n-1} \; \K \; q_{AB} =  
{1\over F(\phi)} \; \left(S_{AB} - {1\over n-1} \; S \; q_{AB}\right),   
\label{E:j-tracefree}  
\end{equation}  
and  
\begin{equation}  
(n-2) \; F(\phi) \; \K - (n-1) \; F'_i(\phi) \; J^i =  - S.  
\label{E:j3}  
\end{equation}  
Inverting the general equations (\ref{E:j2}) and (\ref{E:j3}) yields  
our final form of the generalized junction conditions.  For the trace  
of extrinsic curvature  
\begin{equation}  
\K=  
-{  
S + (n-1) H^{ij}(\phi) \; F'_i(\phi) \;(\L_\brane)'_j  
\over  
(n-2) F(\phi) + (n-1) H^{kl}(\phi) \; F'_k(\phi) \; F'_l(\phi)  
}.  
\label{E:j-trace}  
\end{equation}  
For the normal discontinuity in the scalar derivative  
\begin{equation}  
J^i=  
H^{ij}(\phi)   
\left[  
F'_j \;  
{  
S + (n-1) H^{pq}(\phi) \; F'_p(\phi) \;(\L_\brane)'_q  
\over  
(n-2) F(\phi) + (n-1) H^{kl}(\phi) \; F'_k(\phi) \; F'_l(\phi)  
} 
-  (\L_\brane)'_j  
\right]. 
\label{E:j-phi}  
\end{equation}  
By looking at the expressions (\ref{E:j-tracefree}),  
(\ref{E:j-trace}), and (\ref{E:j-phi}) the first thing that we can  
notice is that for $F(\phi)=1$ neither does $\K_{AB}$ depend on $\L'_b  
$ nor does $J$ depend on $S_{AB}$.  As we have already seen in the  
previous subsection, in the Einstein frame the coupling of the brane  
degrees of freedom to the scalar field and to the metric separately  
induce uncorrelated kinks in the geometry and in the scalar  
field. When $F(\phi)\neq 1 $, however, both effects become  
interconnected: a kink in the scalar field configuration can be  
produced even though there is no direct coupling between the bulk  
scalar field and the fields on the brane, and {\em vice-versa}.  
  
These complete sets of generalized junction conditions [either  
(\ref{E:j1}) and (\ref{E:j2}), or (\ref{E:j-tracefree}),  
(\ref{E:j-trace}), and (\ref{E:j-phi})] are very powerful.  They cover  
most of the scalar fields one can encounter in the literature, and  
have wide applicability beyond the confines of the brane world  
scenario. The most important differences between these junction  
conditions and the standard ones show up when $F(\phi)\neq 1$. In the  
next section we will discuss the peculiarities of these conditions  
when $F(\phi)\neq 1$, that is, when there is a scalar field that is  
not minimally coupled to the geometry.  
  
Finally for completeness we mention that it is still possible (but  
perhaps algebraically not so useful) to reassemble the trace and  
trace-free parts of the extrinsic curvature to yield  
\begin{equation}  
\K_{AB}  
=  
{S_{AB} \over F(\phi) }  
-  
\left\{  
{  
[F(\phi) + H^{ij}(\phi) \; F'_i(\phi)' \; F'_j(\phi) ] S   
+ F(\phi)  \; H^{ij}(\phi)\; F'_i(\phi)\; (\L_\brane)'_j  
\over  
(n-2) F(\phi) +  (n-1) H^{ij}(\phi)\; F'_i(\phi)\;F'_j(\phi)  
}   
\right\}  
\; {q_{AB}\over F(\phi)}.  
\label{E:j-reassemble}  
\end{equation}

\section{The exceptional case}  
\label{S:exception}  

An exceptional case occurs when the denominator in the generalized  
junction conditions vanishes, that is, when  
\begin{equation}  
F(\phi)=  
- {(n-1)\over(n-2)} H^{ij}(\phi) \; F'_i(\phi) \; F'_j(\phi).    
\label{E:kinetic-condition}  
\end{equation}  
This can only occur for $F(\phi) \; H_{ij}(\phi)$ not positive definite  
as we are assuming $n\geq 4$ to avoid too simple a gravity law. Also  
note that a dilaton field never satisfies this exceptionality  
condition, either in the Einstein or the string frames.  
  
In this case the expressions (\ref{E:j-trace}), (\ref{E:j-phi}), and  
(\ref{E:j-reassemble}) are not well defined. However, what this  
condition really means is that the matrix inversion used to go from  
(\ref{E:j2}) and (\ref{E:j3}) to (\ref{E:j-trace}) and (\ref{E:j-phi})  
is singular, though the trace-free relationship (\ref{E:j-tracefree})  
remains perfectly valid. The attempt to invert the singular matrix now  
instead results in a compatibility condition between $S$  and $\L_\brane'$  
\begin{equation}  
(\L_\bulk)'_i = {1\over n-2} \; {F'_i(\phi)\over F(\phi)} \; S,   
\end{equation}  
and a reduced set of linear constraints between $\K$ and $J^i$ [say equations  
(\ref{E:j2})], so that we could view $\K$ (for instance) as freely  
specifiable.

As a particular example, for these exceptional systems one can have 
``phantom branes'' or ``Cheshire cat'' configurations with both a kink 
in the geometry and in the scalar field without the need to add any 
external thin shell of energy.  {From} (\ref{E:j1}) and (\ref{E:j2}), 
taking $\L_\brane(g_{AB},\phi,\psi)\equiv0$ so that both $S_{AB}=0$ 
and $(\L_\brane)'_i=0$, and imposing the exceptionality condition 
(\ref{E:kinetic-condition}) one arrives at 
\begin{equation}  
\K_{AB}=  {1\over n-2} \; {F'_i(\phi) \; J^i\over F(\phi)} \; q_{AB}, 
\qquad 
\hbox{and} 
\qquad 
J^i = - H^{ij}(\phi) \; F'_j(\phi) \; \K, 
\end{equation}  
that is, a simple relation between the jumps in the extrinsic 
curvature and in the normal derivative of the scalar field. (And note 
in particular that there is a unique direction in field space for 
which there is a discontinuity; effectively, in these ``Cheshire cat'' 
configurations only one of the scalar fields is allowed to have a 
discontinuity.) 
  
To better understand what this system means we can perform a conformal  
transformation to the Einstein frame. The above condition between  
$F(\phi)$ and $H(\phi)$, equation (\ref{E:kinetic-condition}), is  
nothing more than the condition to (at the point in question)  
eliminate the kinetic term for {\em one} of the scalar fields in the  
Einstein frame.  (And so locally turn that scalar field equation, in  
the Einstein frame, into an algebraic constraint rather than a  
differential relation.) To see this, note that transforming from the  
generic $F(\phi)\neq1$ frame to the Einstein frame implies the  
conformal redefinition  
\begin{equation}  
g_{AB} = \Omega^2(\phi) \; [g_E]_{AB},  
\end{equation}  
with  
\begin{equation}  
\Omega(\phi)^{n-2} = F(\phi)^{-1}, 
\end{equation}  
and the concomitant redefinition  
\begin{equation}  
H_{ij}(\phi) \to [H_E]_{ij}(\phi)   
= H_{ij}(\phi) + {(n-1)\over(n-2)} { F'_i(\phi) \; F'_j(\phi)\over F(\phi)}.  
\end{equation}  
Then  
\begin{equation}  
\det[H_E] = \det(H)   
\left\{   
1 + {(n-1)\over(n-2)} {H^{ij}(\phi) \; F'_i(\phi) \; F'_j(\phi)\over F(\phi)}   
\right\}.    
\end{equation}  
So $\det[H_E]=0$ is equivalent to the exceptionality condition  
(\ref{E:kinetic-condition}).  
  
If we insist that the exceptionality condition hold as an identity 
throughout field space rather than ``accidentally'' at some specific 
value of the field, then we can think of this kind of scalar field as 
nothing more than the conformal part of the metric in a pure-gravity 
theory (with its corresponding ``negative'' kinetic energy). That is, 
if we start with the ordinary Einstein--Hilbert action and spilt the 
metric into a conformal factor times some fiducial metric, then the 
conformal factor (when viewed as a scalar field propagating in the 
geometry of the fiducial metric), identically satisfies the 
exceptionality condition above. 
 
Indeed if $g_{AB}$ and $\phi$ are the {\em only} fields present in the 
bulk then these ``Cheshire cat'' branes can always be smoothed away in 
this manner, leaving absolutely no trace of their existence, and so 
they should then be viewed as completely unphysical mathematical 
artifacts. If there are several scalar fields, then this argument still 
goes through (modulo potential problems from globally diagonalizing 
the Einstein frame sigma model metric $[H_E]_{ij}$). 
 
On the other hand, consider the very intriguing possibility that the 
exceptionality condition holds not as an identity in field space, but 
rather as an ``accident'' at a point in field space corresponding to 
the value of the scalar fields on the brane.  Then ``Cheshire cat'' 
branes of this type cannot be globally transformed away in the above 
manner. You can eliminate the kink in the geometry, and the kink in 
the scalar fields, but there will now be at least one scalar field 
with a very peculiar kinetic energy term: a kinetic energy that 
vanishes at some (but not all) points in field space.  There is a 
conservation of difficulty and one has a tradeoff between kinks in the 
geometry and scalar fields {\em versus} exceedingly peculiar kinetic 
energies. 
  
\section{Moduli fields and brane tensions}  
\label{S:moduli}  

Let us now restrict the discussion to ``bare'' brane Lagrangians of  
the form  
\begin{equation}  
\L_\brane= - T \; f(\phi),  
\end{equation}  
that is, the brane only provides a vacuum energy modulated by the   
value of the scalar field:   
\begin{equation}  
S_{AB}=- T \; f(\phi) \; q_{AB}.  
\end{equation}  
An immediate consequence is that the trace-free part of $\K_{AB}$ is  
zero,   
\begin{equation}  
\K_{AB} = {1\over n-1} \; K \; q_{AB},  
\end{equation}  
and that all of the interesting physics is hiding in the trace $\K$  
and in $J$.  
  
\subsection{Dilaton field in the Einstein frame}  
  
Consider $F(\phi)=1$ and $H(\phi)=4/(n-2)$. This is the common case  
analyzed in the literature --- a dilaton field in the Einstein  
frame~\cite{lambdaprob}. Then a positive brane tension, $T  
\;f(\phi)>0$ yields a warp factor that decreases when going away from  
the brane.  
(Strictly speaking this is true only when one imposes a $Z_2$ symmetry  
on the solutions, or adopts the ``one-sided'' view discussed  
in~\cite{barcelo,barcelo-brane}. More generally we have to say that is the  
sum of the variation of the warp factor when departing from the brane  
in the two possible directions that decreases away from the brane.)  
  
The behaviour can be easily read from (\ref{E:israel-trace}) and  
(\ref{E:j-einstein}) [and also from the more general (\ref{E:j-trace})  
and (\ref{E:j-phi}) when particularized to this case]:  
\begin{equation}  
\K=- {1\over n-2} \; S = {n-1\over n-2} \; T \; f(\phi).   
\end{equation}  
The behaviour of the scalar field away from the brane depends upon  
the derivative of $f(\phi)$,  
\begin{equation}  
J = - {n-2\over 4} \; T\; f'(\phi).  
\end{equation}  
For positive tension branes ($T\;f(\phi)>0$), the condition $[\ln  
f(\phi)]'>0$ yields a scalar field that decreases away from the brane;  
with $[\ln f(\phi)]'<0$ the scalar field increases. Generically it is  
found that positive tension branes give rise to bulk configurations in  
which at some finite distance from the brane the spacetime  
``terminates'' in the form of a naked singularity~\cite{lambdaprob}.  
  
When $F(\phi)\neq 1$ this is no longer true. The existence of a  
coupling between the scalar field and the curvature makes the junction  
conditions more involved. The thin shell of energy contributes not  
only to the kink in the geometry but also to that in the scalar field.  
To see this with more detail, let us consider the case of a dilaton  
field in the string frame in $5$ dimensions.  
  
\subsection{Dilaton field in the string frame}  
  
Consider $F(\phi)=\exp({-2\phi})$ and $H(\phi)=-4\; \exp({-2\phi})$.  
Write the junction conditions (\ref{E:j-trace}) and (\ref{E:j-phi}) as  
\begin{equation}  
\K= - (n-1) \; T \;  e^{2\phi}  
\left[f(\phi)+{1 \over 2} f'(\phi)\right],  
\end{equation}  
\begin{equation}  
J= T \; e^{2\phi} \;   
\left[  
{n-1\over2} f(\phi)+{n-2\over 4} f'(\phi)  
\right].  
\end{equation}  
Imagine that we take an ansatz for the scalar field-brane coupling  
function $f(\phi)=e^{-\alpha\phi}$. Then, for  
$2(n-1)/(n-2)<\alpha<+\infty$ the warp factor and the dilaton field  
decrease away from the brane.  For $\alpha=2(n-1)/(n-2)$ the warp  
factor still decreases away from the brane but the dilaton  
configuration has no kink.  For $2<\alpha<2(n-1)/(n-2)$, the warp  
factor decreases away from the brane but the dilaton field increases.  
For $\alpha=2$, which is the case usually chosen as the appropriate  
coupling, there is no kink in the geometry and the dilaton increases  
away from the brane. For $\alpha<2$ (the case $f(\phi)=1$ would  
correspond to $\alpha=0$) the warp factor and the dilaton increase  
away from the brane.  
  
In view of the previous discussion some observations are in order. Let 
us take a coupling function $f(\phi)=e^{-2\phi}$, and restrict to the 
case $n=5$. This is used commonly as representing the coupling of the 
dilaton field to the brane~\cite{lambdaprob}; (in the Einstein frame 
used in those works the corresponding coupling function is 
$f_{E}(\phi)=e^{2\phi/3}$).  For this case we see that we can have 
3-branes with arbitrarily large tensions but, nonetheless, with a 
smooth string-frame geometry (that is, without any hypersurface on 
which there is a jump in the extrinsic curvature).  We would notice 
the existence of such a 3-brane only through its effect on the scalar 
field configuration.  However, in transforming this metric-scalar 
configuration to the Einstein frame we will recover a kink in the 
Einstein frame geometry at the location of the 3-brane. This is due to 
the fact that the Einstein frame metric is a function of both the 
string frame metric and the scalar field. From the point of view of 
the Einstein frame is the existence of a kink in the geometry that is 
trapping 4-dimensional gravitons near the brane world. From the string 
frame point of view, this trapping will be due to a non-differentiable 
variation of the effective Newton constant when moving across the 
brane. From the Einstein frame point of view a finite 4-dimensional 
Newton constant would show up in the brane if the extra dimension has a 
finite volume per unit area. 
(By this we mean that $\int a(\eta) \;\d\eta = < \infty $ where 
$a(\eta)$ is the scale factor of the geometry.) 
Taking the string frame point of view, the relevant quantity is the 
volume of the extra dimension weighted by the scale factor derived 
from the effective Newton constant.

\section{Summary and discussion}  
\label{S:summary}  

We have generalized the Israel--Lanczos--Sen thin-shell junction 
conditions to the case in which the bulk spacetime is filled, apart 
form gravity itself, by a set of general scalar fields with 
non-trivial curvature coupling. First, we have derived the standard 
Israel--Lanczos--Sen junction conditions beginning directly from the 
gravitational action. The method we have followed turns out to be 
straightforwardly generalizable in the presence of other fields in the 
bulk. We have found these new junction conditions for gravity plus a 
very general set of moduli fields with arbitrary couplings to the 
scalar curvature. Our formalism is capable of dealing with the dilaton 
field, in the string and/or Einstein frames, metric-scalar 
theories {\it \`a la} Brans--Dicke, and non-minimally coupled scalars. 
 
The existence of couplings between the moduli fields and the curvature 
scalar makes the generalized junction conditions interconnected.  One 
can have a ``kink'' in the scalar field configuration even though 
there is no direct coupling between the scalar fields and the thin 
shell of stress-energy. Conversely, these couplings between the 
thin-shell and the scalar fields make contributions to the form and 
strength of the ``kink'' in the geometry. 
 
An exceptional case appears for some moduli field configurations. 
These exceptional field configurations are characterized by the 
vanishing of the kinetic term of some of the scalar fields when 
transforming to the Einstein frame. In this exceptional case we 
showed that one can find ``kink'' metric-scalar configurations in the 
original frame even in the absence of any thin-shell of 
stress-energy. The ``kink'' in the geometry and that in the scalar 
fields feed back into each other. The possibility that the exceptional 
condition could be satisfied for some particular values of the moduli 
fields could give place to very interesting ``Cheshire cat'' 
configurations. 
 
Particularizing to the dilaton field, we showed that the geometric 
picture that emerges when working in the string or Einstein frames are 
very different. It is even possible that a geometry with a ``kink'' in 
the Einstein frame has a counterpart which in the string frame is 
perfectly smooth. The Einstein frame geometric ``kink'' is absorbed 
into the conformal factor in such a way that in the string frame only 
the dilaton field undergoes a discontinuity in its normal derivative 
as one crosses the brane. These apparently different configurations 
are however equivalent. Up to this point in the discussion the choice 
of frame is a matter of convenience. 
 
However this might not always be the case.  First, the complete 
equivalence between frames is only guaranteed if the conformal 
transformation relating them is not singular. When this is not true 
one can find configurations in one frame that have no complete 
analogue in the other frame. As an specific example, for a 
non-minimally coupled scalar field the present authors have found some 
traversable wormhole solutions~\cite{scalars} for which the term 
$\kappa - \xi \phi^2$ becomes zero in some specific points of the 
geometry (on the other hand, the ``Jordan'' frame geometry is 
perfectly regular). The relation between de Einstein and ``Jordan'' 
frames can be written as $[g_E]_{AB} =(\kappa - \xi 
\phi^2)[g_J]_{AB}$.  Therefore, the whole wormhole configuration in 
the ``Jordan'' frame has no non-singular counterpart in the Einstein 
frame. 
 
Second, complete equivalence between choice of conformal frames is 
guaranteed only if you simultaneously change both frame {\em and the 
equation of motion for test particles}. If physical test particles are 
observed to follow geodesics in one conformal frame then they will not 
follow geodesics in any other conformal frame. If free fall is truly 
universal (as the E\"otv\"os experiments indicate), then one conformal 
frame is more equal than the others. (Brans--Dicke type theories are 
set up in such a manner that this special frame is the Jordan frame, 
and for this reason the Jordan frame is often called the ``physical'' 
frame.)  Thus, by performing experiments on bulk matter one should be 
able to distinguish between different frames~\cite{casadio}, and the 
discussion about which frame is the ``most physical'' will come into 
play~\cite{faraoni,magnano}. We emphasise the well-known, albeit 
commonly overlooked fact that string theory does not satisfy the 
weak equivalence principle, (except approximately), let alone the 
Einstein equivalence principle~\cite{casadio,Polchinski,Will}.  If one 
uses experiment to decide which is the ``most physical'' frame, then 
to work in another frame would be nothing more than a mathematical 
trick (although sometimes potentially useful). 
 
However, taking the Randall-Sundrum models as background scenario it 
is far from clear if the differences in bulk physics will affect 
experiments performed in the brane itself. Since bulk matter is now 
trapped on the brane, direct free-fall experiments could at best 
suggest the choice of a particular frame on the brane, but would leave 
the off-brane conformal frame a free variable whose choice would be 
driven by esthetics and mathematical convenience rather than by 
physics. 
 
At a more basic level, when proposing the fundamental Lagrangian, for 
each new field that one adds it is necessary to decide in which of the 
previous possible frames it will couple minimally.  For example, if we 
want to begin with a non-zero cosmological constant for the bulk, 
defined as a proper tendency of the bulk spacetime to be curved, in 
which frame should the expression $G_{AB}=-\Lambda g_{AB}$ hold? 
Side-stepping the answer to this question for now, it is imperative to 
analyze (as in this paper) systems with very general scalar 
potentials, to obtain results that can be applied in any frame.

\section*{Acknowledgments}  
 
The research of CB was supported by the Spanish Ministry of Education  
and Culture (MEC). MV was supported by the US Department of Energy.

\clearpage  
   
\end{document}